\documentclass[11pt,twoside]{article}

\usepackage{asp2006}
\usepackage{epsf}
\usepackage{psfig}
\usepackage{lscape}

\begin{document}
\title{Light Curves of Galactic Eclipsing\\ Double Periodic Variables}   
\author{G. Michalska$^1$, R.E. Mennickent$^1$, Z. Ko{\l}aczkowski$^{1,2}$, G. Djura\v{s}evi\'{c}$^3$}
\affil{\scriptsize
$^1$Universidad de Concepci\'on, Departamento de Astronom\'{\i}a, Casilla 160-C, Concepci\'{o}n, Chile (gabi@astro-udec.cl,rmennick@astro-udec.cl)\\
$^2$Instytut Astronomiczny, Uniwersytet Wroc{\l}awski, Kopernika 11, 51-622 Wroc{\l}aw, Poland\\
$^3$Astronomical Observatory, Volgina 7, 11160 Belgrade, Serbia}
\begin{abstract} 
We present results of the investigation of the nature of double periodic variables (DPVs). We have selected a sample of Galactic eclipsing DPVs for a multiwavelength photometric study aimed to reveal their nature. The short orbital periodicity and the cyclic variability are decoupled and separately investigated. Shapes of orbital light curves are consistent with semi-detached binaries. The amplitude of the long cycle is always larger in redder bandpasses.  
\end{abstract}

\section{Introduction}
Double Periodic Variables (DPVs) is a group of binary stars with orbital period between 1 and 16~d, characterized by additional long cyclic variability in the range of 50-600~d, correlated with the orbital period (Mennickent et al.~2003, Mennickent \& Ko{\l}aczkowski 2009a). These stars were discovered during the search for Be stars in the Magellanic Clouds based on the OGLE variable stars catalogue. A careful survey of the OGLE and MACHO photometry, allowed to detect this kind of variability in about a hundred stars in the Magellanic Clouds. The first results obtained for one LMC DPV star, OGLE05155332-6925581, were recently published (Mennickent et al.~2008).  Investigating the ASAS photometric data we have found eleven Galactic DPVs. The light curves of four Galactic DPVs are shown in Fig.1. After the discovery of this new group of variable stars, we have initiated a multiwavelength observing campaign.

\section{Observations and analysis}

The photometric observations of selected DPVs were carried out with the 0.6-m REM telescope located at La Silla Observatory. From February 2008 we have taken above 320 frames in V and I filters and 160 frames in J and K filters for every star. In order to separate the short- and long-term variability, we used the Fourier decomposition technique. The four-colour light curves of one DPV presented in Fig.2. show that the amplitude of long-term variability in V filter is the lowest, while in K filter is the largest. This is a common feature of all DPVs.
The analysis of high-resolution spectroscopy allowed to detect at least three spectral components: (1) absorption lines typical for B-A type stars, (2) an optically thin H$\alpha$ emitting region, (3) an optically thick region forming variable and broad helium absorption lines.
The DPV light curve are being modelled by applying the inverse-problem method and the code described in Djura\v{s}evi\'{c} et al.~(2008). For V393 Sco we assumed mass ratio 0.41, temperature of the less massive star, filling its Roche lobe, equal to 7900 K and the presence of two active bright regions on the disc edge. As result we obtained the temperature of the more massive star equal to 17400 K, and masses equal to 9.9 and 4.1M$_\odot$.

\begin{figure}[!ht]
\centerline{\hbox{\psfig{figure=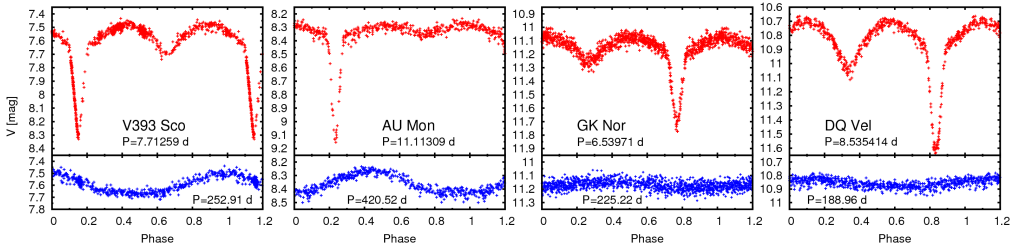,angle=0,clip=,height=3.25cm}}}
\caption{Phase diagrams of the galactic DPV after the Fourier decomposition of ASAS data ({\it upper:} orbital variability, {\it lower:} long variability).}
\label{fig1}
\end{figure}
\begin{figure}[!ht]
\centerline{\hbox{\psfig{figure=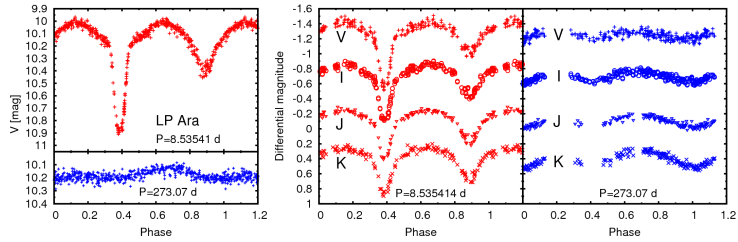,angle=0,clip=,height=3.25cm}}}
\caption{{\it Left:} the same as in Fig.1. {\it Right:} orbital and long variability found in REM data. Note the larger amplitude in the K band.}
\label{fig2}
\end{figure}

\section{Conclusions}

The analysis of the available data, suggests that DPVs are binaries in a mass exchange evolution stage. Probably the primary is rotating at the critical velocity and matter cumulates in a circumprimary disc. Disc size increases until certain instability ejects mass outside the binary system. As a result, in DPVs we observe the long-term variability (Mennickent \& Ko{\l}aczkowski 2009abc).



\acknowledgements 
This work was supported by Chilean Fondecyt Projects 3085010 and
 1070705 and the Serbian Ministry of Science Project 146003.

\end{document}